\documentclass[fleqn,twoside]{article}
\usepackage{espcrc2}

\usepackage{epsfig}

\usepackage{graphicx}
\usepackage[figuresright]{rotating}

\newcommand{\etal}{{\it et al.}}
\newcommand{\AmS}{{\protect\the\textfont2
  A\kern-.1667em\lower.5ex\hbox{M}\kern-.125emS}}

\title{Multianode Photo Multiplier Tubes as Photo Detectors for Ring Imaging 
Cherenkov Detectors}

\author{F. Muheim\address[edin]{Department of Physics and Astronomy,
	University of Edinburgh \\
        Mayfield Road, Edinburgh EH9 3JZ, Scotland/UK}
}
\begin{document}

\begin{abstract}

The  64-channel Multianode Photo Multiplier (MaPMT) has been evaluated 
as a candidate for the LHCb Ring Imaging Cherenkov (RICH) 
photo detectors.  We present result from data taken with a 3x3 array 
of closely packed MaPMTs mounted onto the RICH 1 prototype vessel,
exposed to  charged particle beams at CERN,
and read out at LHC speed. 
Using a LED light source, 
we have performed spatial light scans to study 
the light collection efficiency of the MaPMTs 
We have also measured the performance of 
the MaPMTs as a function of the applied high voltage. Different dynode 
resistor chains have been used to study the tubes at low gains. 
In addition, we have studied the behaviour of the MaPMT in magnetic fields.

\vspace{1pc}
\end{abstract}

\maketitle

\section{Introduction}

We have evaluated 
the  64-channel Multianode Photo Multiplier (MaPMT) 
as a candidate photo detector for the LHCb Ring Imaging Cherenkov (RICH) 
counters. 
The MaPMT contains a 8x8 array of 64 dynode chains enclosed in a single vacuum envelope. 
The window is made of UV glass 
instead of borosilicate which extends the transparency down to 200 nm.  
According to the manufacturer, the bi-alkali photo cathode has a quantum
efficiency of 22\% at a wavelength of 380~nm. 
The pitch between anode pixels is 2.3 mm.
When accounting for the inactive 0.2 mm gaps between pixels,
the MaPMT has an active area coverage of only 38\%.
The full angular coverage with 85\% active area 
can be 
restored by mounting a fused silica 
lens with one flat and one spherical surface
with a radius of curvature of 25 mm 
in front of the tube. The lens has a
demagnification factor of about two-third
and focuses the incoming photons onto the sensitive area of the MaPMT.

\section{Test Beam Results}

The sensitivity of the MaPMT to Cherenkov photons
has been tested with particle beams of 120 GeV/c pions 
at the SPS accelerator at CERN. 
A detailed account of these
tests is described in reference~\cite{Albr02}.
Here we present the main results. 
A 3x3 cluster of closely packed 
MaPMTs has been mounted onto a prototype of the RICH\,1
detector for the LHCb experiment. 
The radiator gas was CF$_4$ at a pressure of 700 mbar
and the photo cathode was set at -1000 V. 
To demonstrate that the MaPMTs will work at the LHC  
the data were recorded with the APVm chip running at 40 MHz. 

\begin{figure}[bht]
\begin{center}
\mbox{\epsfig{file=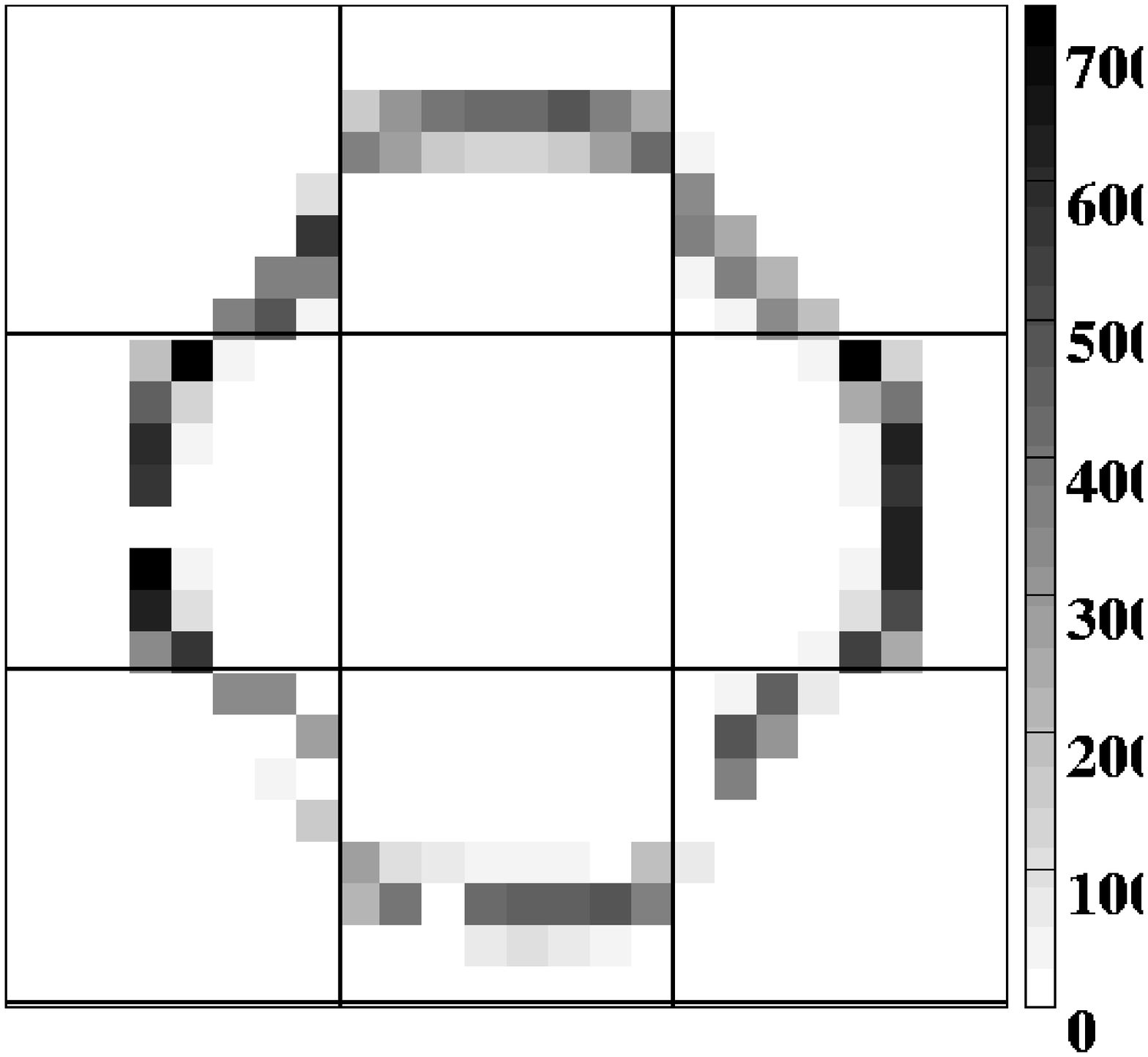,width=0.49\linewidth}
	\epsfig{file=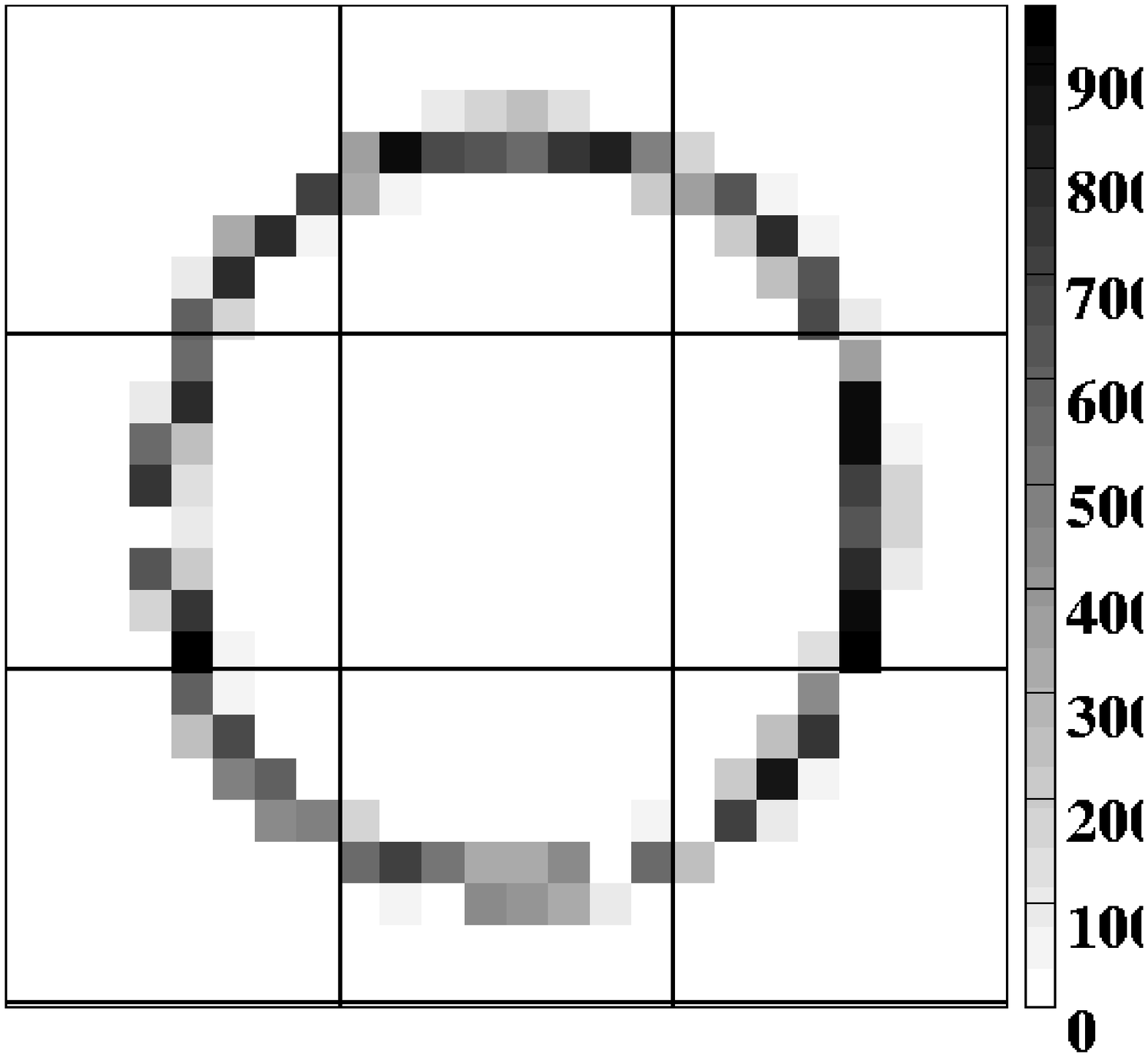,width=0.49\linewidth}}
\vspace*{-5mm}
\caption{Cherenkov ring measured with the MaPMT array with (right plot)
and without lenses (left plot) mounted in front of the tubes.}
\label{fig:comparison}
\end{center}
\end{figure}
In Figure~\ref{fig:comparison} 
we show the results for two runs of 6000 events each, 
one with and one without the lenses mounted in front of the MaPMTs.
In both cases, a ring of Cherenkov photons is clearly visible.
The ratio of yields with and without lenses is 1.55
as is expected from a simulation. 
With the lenses the full Cherenkov ring is captured
which demonstrates that MaPMTs 
can be closely packed to restore full geometrical acceptance. 
We observe $6.96 \pm 0.33$ photo electrons per event,
in very good agreement with the simulation. 
The corresponding figure of merit 
for a RICH detector is $N_0 = 110 \pm 7 \; \rm cm^{-1}$.  
\begin{figure}[htb]
\begin{center}
\mbox{\epsfig{file=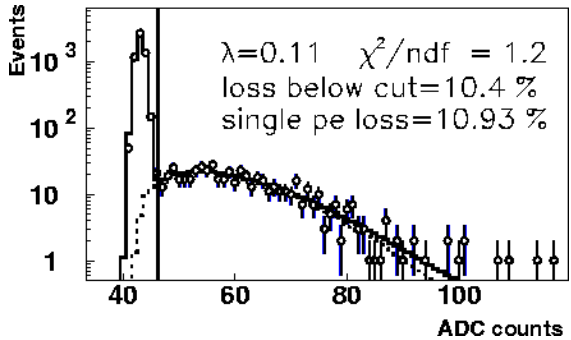,width=0.99\linewidth}}
\vspace*{-5mm}
\caption{MaPMT pulse height spectrum with
a superimposed fit (solid line). Also shown is the
single photo electron component (dashed line). The vertical line 
indicates the $5\sigma$ cut above pedestal.}
\label{fig:pulse-height}
\end{center}
\end{figure}
In Figure~\ref{fig:pulse-height} 
we show the pulse height spectrum for a single pixel.
The data were corrected as follows. 
A  common-mode baseline has been subtracted and signals due to cross-talk
stemming from the read-out electronics have been removed.
We clearly distinguish between the pedestal peak
and the signal containing mainly one photo electron.
Overlaid is a fit where 
$\lambda$ is the mean number of photo electrons.

\section{LED Scans, Signal Shape, Low Gains}

We have also studied the performance of MaPMTs 
in the laboratory. As a light source we used 
a blue LED
coupled to a mono-mode fibre which could be
moved with respect to the MaPMT by semi-automated optical stages.
Two read-out systems were employed. One is 
the 40 MHz read-out with the APVm chip
also used for the beam tests, and another one is based on  
CAMAC amplifiers and ADCs.

In addition to the nine tubes studied with test beams
we obtained two more MaPMTs which,
according to the manufacturer, have an increased quantum efficiency
of maximum 25 - 27\% at 360~nm. The focusing of these new tubes has 
been improved. Firstly, the acceptance at the outside edges
of the active area of the MaPMT has been increased,
and another focusing wire has been added.
Secondly, the distance
between the focusing grid and the entry slits to the dynode
chain has been reduced. 
We have studied these new MaPMTs 
with a LED light scan. 
The data have been taken by moving the light source 
in steps of 0.1 mm 
along the front face of the MaPMT,
with the photocathode at -900 V. 
\begin{figure}[htb]
\begin{center}
\mbox{\epsfig{file=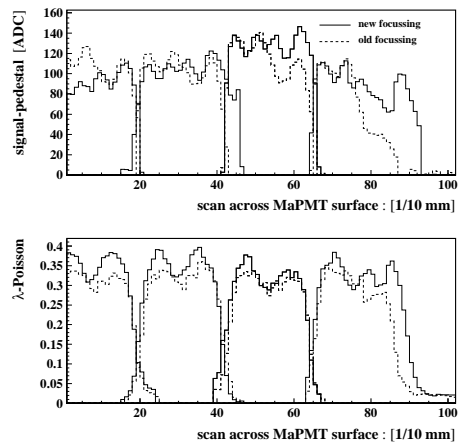,width=0.9\linewidth}}
\caption{Scan across an MaPMT. Shown are the 
average single photon pulse height $s$ (top) and the 
mean number of photo electrons $\lambda$ (bottom) 
versus the relative position.}
\label{fig:linscan}
\end{center}
\end{figure}
The pulse height spectra have been fit with a Gaussian curve as 
single photo electron signal shape.
Shown in Figure~\ref{fig:linscan}  
are the average single photon pulse height, $s$,  and 
$\lambda$, the mean number of measured photo electrons
as a function of the relative position along the tube.
The gain is proportional to $s$ and the collection efficiency  
is closely related to $\lambda$. 
We observe that the edge pixels have a larger acceptance
for the new MaPMT (solid histograms) 
with respect to the standard focusing tube (dashed histograms).
Due to the improved focusing, 
both the gain and  the collection efficiency  
are more homogeneous. 

Most front-end chips developed for the LHC have been
optimised for signals from Silicon sensors or 
micro-strip gas chambers with a charge of 26000 - 40000~$e$.  
At a voltage of 800 V between anode and photocathode
the gain of the MaPMT is around 300000 which is
about 10 times higher.
To make use of one of these front-end chips for the
the MaPMT a  gain adaptation is needed.
The preferred solution is to modify a preamplifier to a lower gain.
A second possibility is to make use of an attenuator network.
This has been used for the APVm read-out by means of an AC coupler.
However this approach
suffers from cross talk problems. 
A third way that we have investigated is the feasibility to run
the MaPMTs at a lower gain. 

To reduce the gain of the MaPMT we have changed the values 
of the  resistors between the photo cathode, 
the first, second and the third dynode, respectively.
The basic idea is to keep the gain at the first dynode 
while lowering the overall gain.
The values for the default, a medium and a low gain option
are given in Table~\ref{tab:dynode}.
\begin{table*}[htb]
\caption{Resistances in $ \left[ 10^2 \; \rm k \Omega \right] $ in front of each dynode.}
\label{tab:dynode}
\newcommand{\m}{\hphantom{$-$}}
\newcommand{\cc}[1]{\multicolumn{1}{c}{#1}}
\renewcommand{\arraystretch}{1.2} % enlarge line spacing
\begin{tabular}{@{}llllllllllllll}
%\begin{tabular}{@{}lllllllll}
\hline
Dynode        	& 1 & 2 & 3 & 4 & 5 & 6 & 7 & 8 & 9 & 10 & 11 & 12 & Anode \\
\hline
Standard gain 	& 3 & 2 & 2 & 1 & 1 & 1 & 1 & 1 & 1 & 1  & 1  & 2  & 5 \\
Medium gain 	& 4 & 2 & 2 & 1 & 1 & 1 & 1 & 1 & 1 & 1  & 1  & 2  & 5 \\
Low gain 	& 4 & 3 & 3 & 1 & 1 & 1 & 1 & 1 & 1 & 1  & 1  & 2  & 5 \\
\hline
\end{tabular}\\[2pt]
\end{table*}
The signal width is mainly due to the Poisson distribution
for multiplication at the first dynode and can be estimated
as $g_1 = s^2/\sigma^2$
where $\sigma$ is the Gaussian width of the  single photon signal. 
Using the CAMAC read-out
we have measured the overall gain of the MaPMT 
for a center and a border pixel
as a function of
the high voltage for the three different resistor chains.
\begin{figure}[htb]
\begin{center}
\mbox{\epsfig{file=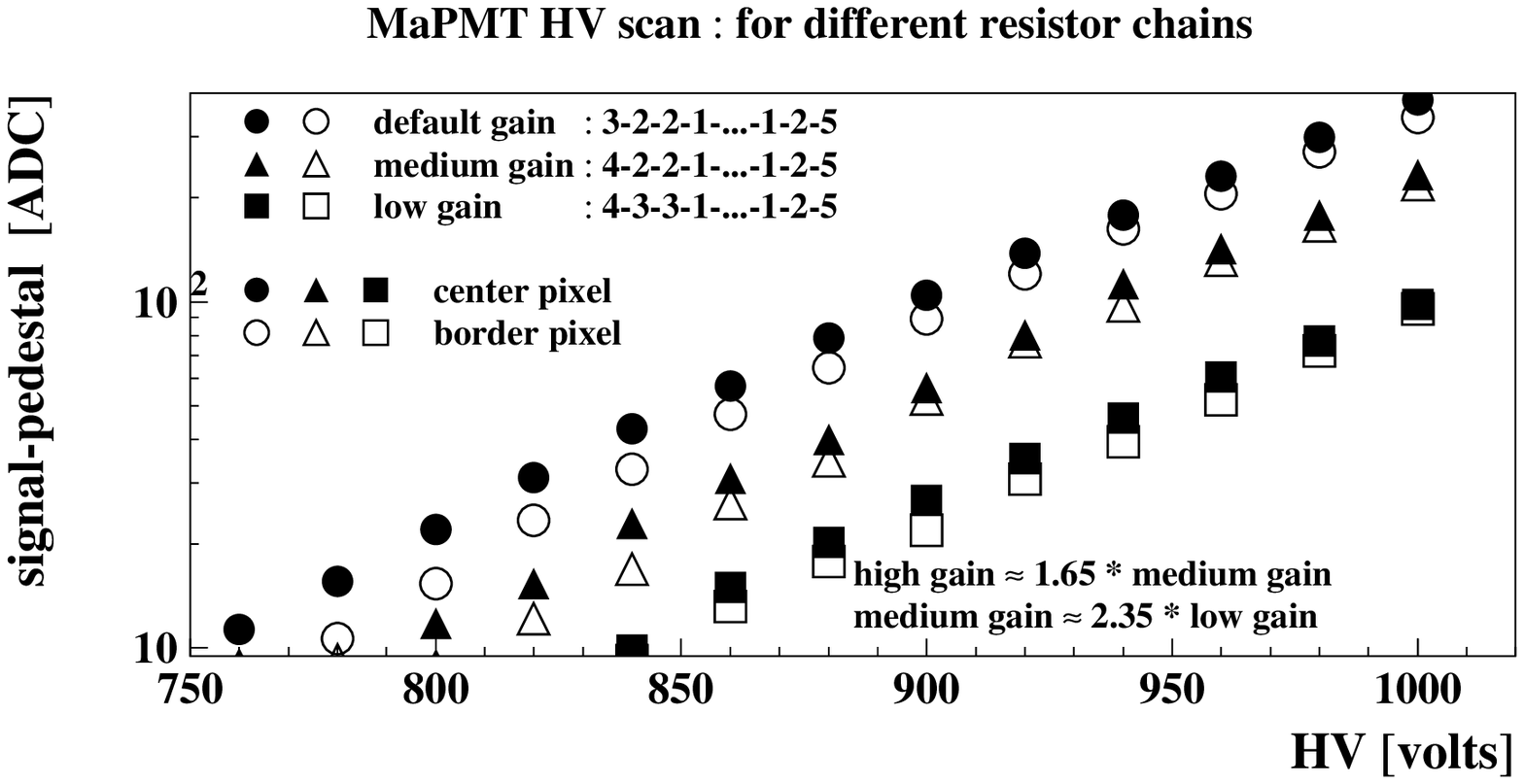,width=0.9\linewidth}}
\mbox{\epsfig{file=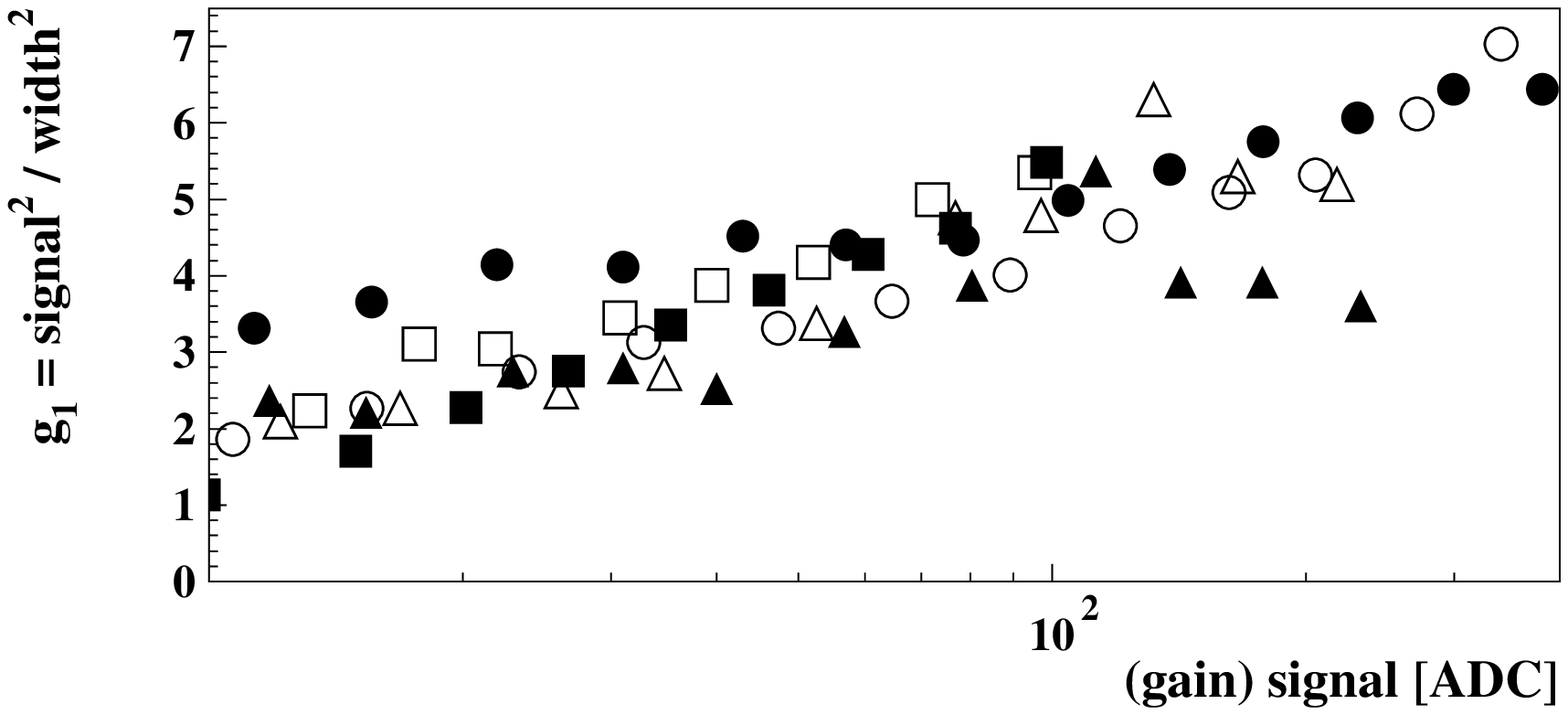,width=0.9\linewidth}}
\caption{High voltage scan for three different resistor chains. 
Top)  Average single photon pulse height $s$ versus high voltage.
Bottom) Gain at the first dynode $g_1 = s^2/\sigma^2$ versus $s$.   }
\label{fig:hvscan}
\end{center}
\end{figure}
In Figure~\ref{fig:hvscan} top) 
we plot $s$ as a function of the applied high voltage 
for the three different resistor chains.
At a given voltage the gain of the MaPMT 
reduces by a factor of four 
for the low gain option.
In the bottom plot we plot 
$g_1 = s^2/\sigma^2$ as a function of $s$. 
As expected we observe that $g_1$  increases with $s$. 
However there are no differences 
between the different resistor chains.
%which  shows that the argument made above is too simple.
We have repeated these measurements with the APVm read-out 
and the results agree.

We have studied the loss of signal below the threshold cut
of $5\; \sigma_{ped}$ where $\sigma_{ped}$ is the Gaussian width
of the pedestal peak. We have employed different
fit methods. Besides a Gaussian signal shape
we have implemented a fit
which uses 
a Poisson distribution for the first two photo electrons~\cite{tokar}.
This method also allows for photons 
which pass through the photo cathode to be converted into photo
electrons at the first dynode.
\begin{figure}[htb]
\begin{center}
\mbox{\epsfig{file=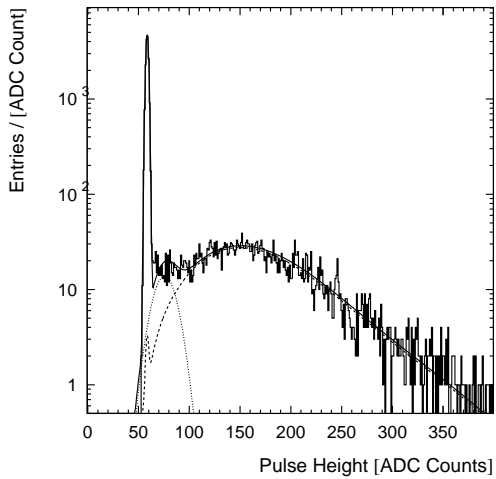,width=0.9\linewidth}}
\caption{MaPMT pulse height spectrum.
The overlaid fit is described in the text. }
\label{fig:poiss-dyn}
\end{center}
\end{figure}
A Gaussian or a Poisson fit to the pulse height spectra
are inadequate to describe 
the region of small signals close to the pedestal. 
In Figure~\ref{fig:poiss-dyn}
we show a pulse height spectra recorded with the CAMAC read-out 
and the photo cathode at -900 V.
The superimposed fit (dotted line) 
allows for photo electrons to be produced at the first dynode.
This fit still does not follow the data points correctly
for small signals. 
There is a broad additional contribution of small pulse
heights, but  
no evidence for a second peak in the data. 
Including a contribution due to 
production of photo
electrons at the first dynode improves the description
of the data.
However, the fit improves the overall description of the data and,
for the spectrum shown above, 
yields a gain at the first dynode of $K_1 = 6.1$ which is in 
agreement with expectations whereas the Gaussian and Poisson fit 
give $K_1 \geq g_1 = 3.2$ and $K_1 = 4.1$, respectively.
The parameter $\lambda$ is independent of the fit method.
Due to gain variations for pixels within a tube 
the loss of signal below the threshold varies from pixel
to pixel. Using the Poisson fit including the ``1st dynode'' effect 
gives the most reliable results. 
The average signal loss of the  64 pixels of an MaPMT is 13\%.  
We have also measured that the signal loss below 
threshold scales with $s$.
Consequently, running at lower gains increases this loss.

\section{Beetle Chip Studies }
The APVm read-out is 
not compatible with the LHCb architecture. 
A possible candidate to read out
the MaPMT is the Beetle chip. 
It can be run in analogue and binary mode.
The binary mode has several advantages, as it ties in better
with the Pixel HPD development for LHCb, described in 
reference~\cite{sajan}, 
which also has a binary read-out,
and for which a design of the Off-Detector Electronics (ODE)
is underway. In addition there would be substantial cost savings.

A program has been undertaken to develop a Beetle adaptation
to the MaPMT. 
By adding a charge and a voltage attenuator at the input
of the preamplifier we have lowered the gain of the Beetle chip.
Test structures
have been produced and studied for this design. 
For the charge attenuator, measurements with a test charge
agree very well with the simulation. 
The dynamic range of the preamplifier is 10 photo electrons.
This preamplifier has been mounted onto a MaPMT and 
a pulse height spectrum from an LED light source has been measured. 
The signal from mostly single photo electrons is nicely separated
from the pedestal.
In this approach the load capacitance affects the gain which would
require to keep the input capacitances of all channels within a few pF. 

\section{Magnetic Field Studies}

A consequence of the optimisation of the LHCb spectrometer~\cite{sajan}
is that now 
the RICH\,1 photo detectors will be placed
 in a region where the magnetic
field strength is about 40~mT. 
We have studied the sensitivity of the
MaPMT to longitudinal and transverse magnetic fields up to $B = 35$~mT.
A set-up with 4 LEDs has been built to 
allow for a diffuse illumination of the MaPMT. The APVm
read-out system was used. 

\begin{figure}[htb]
\begin{center}
%
%\vspace*{-5mm}
\begin{turn}{-90}
\mbox{\epsfig{file=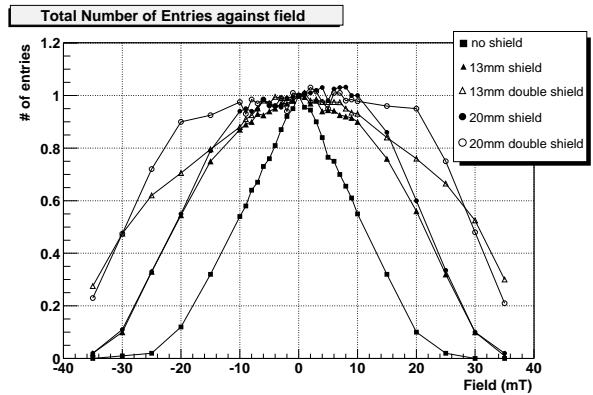,width=0.75\linewidth}}
\end{turn}
%\vspace*{-5mm}
%
\caption{The number of observed photo electrons relative to $B = 0$~mT
versus magnetic field.}
\label{fig:blong}
\end{center}
\end{figure}

We observe that the gain and the collection efficiency of  
a MaPMT decreases with increasing $B$ field. 
At \mbox{$B = 3$~mT},
this effect is already  sizable
for the two edge rows of the tube.
Averaged for all pixels of a tube 
the number of measured photo electrons 
for  $B \geq 3$~mT
is below 90\% of that measured at 0~mT.
The MaPMT can be shielded by enclosing it laterally with 
a \mbox{$\mu$-metal} case. 
This shielding should extend beyond the photo cathode
of the tube.
In Figure~\ref{fig:blong} we show the  
measured number of photo electrons for all pixels
of a MaPMT relative to its value at 0 mT
for  the following set-ups: 
No shielding, 
0.9 mm thick shielding extending by 13 mm and 20 mm,
and 1.8 mm thick shielding extending by 13 mm and 20 mm.
We observe that the shielding of the tube is effective. 
Already for the single 0.9~mm thick shielding the MaPMT functions
in longitudinal magnetic fields up to 10~mT.
We have also studied the behaviour of the MaPMT in
transverse  fields.
These measurements show that the MaPMT is insensitive
to transverse magnetic fields up to 25~mT.

\section{Conclusions}

Using particle beams at the SPS at CERN
we have successfully tested a 3x3 array of 
multianode photomultiplier tubes. We have demonstrated
that by means of lenses mounted 
in front of the closely packed MaPMTs
the Cherenkov photons are focused onto the sensitive area
of the devices.
Using a laboratory with LED light sources we have studied
the performance and the signal shape of the MaPMTs in detail.
A change in the focusing of the photo electrons onto the first dynode 
improved the homogeneity of the gain and of the light collection.
The gain and width of the signal as well the signal loss
have been measured for voltages between $-700$ and $-1000$ V.  
We have demonstrated that running the MaPMT at lower gains
significantly increases the signal loss below
threshold.
Finally, we have evaluated the sensitivity of the MaPMT 
to longitudinal and transverse magnetic fields up to 35~mT.

\section{Acknowledgements}

I thank Reinhardt Chamonal, Stephan Eisenhardt, and Dave Websdale
for the help in preparing this talk. 
This workshop was dedicated to the memory of 
Tom Ypsilantis whom I had the privilege to work with.

\end{document}